# In-plane magnetic field induced double fan spin structure with *c*-axis component in metallic kagome antiferromagnet YMn$_6$Sn$_6$

Kelly J. Neubauer[1,*], Chunruo Duan[1,*], Feng Ye[2], Rui Zhang[1], Songxue Chi[2], Qi Wang[3], Kathryn Krycka[4], Hechang Lei[3], Pengcheng Dai[1,**]

[1]*Department of Physics and Astronomy,*
*Rice University, Houston, Texas 77005, USA*
[2]*Neutron Scattering Division, Oak Ridge National Laboratory,*
*Oak Ridge, Tennessee 37831, USA*
[3]*Department of Physics and Beijing Key Laboratory of Opto-electronic Functional Materials &*
*Micro-nano Devices, Renmin University of China, Beijing 100872, China*
[4]*NIST Center for Neutron Research,*
*National Institute of Standards and Technology, Gaithersburg, MD 20899, USA*
[*]*These authors made equal contributions to the present work*
[**]*email: pdai@rice.edu*
(Dated: July 22, 2020)

**Abstract:**
The geometrical frustration nature of the kagome lattice makes it a great host to flat electronic band, non-trivial topological properties, and novel magnetisms. Metallic kagome antiferromagnet YMn$_6$Sn$_6$ exhibits the topological Hall effect (THE) when an in-plane magnetic field is applied. THE is typically associated with the nanometer-sized non-coplanar spin structure of skyrmions in non-centrosymmetric magnets with large Dzyaloshinskii-Moriya interaction. Here we use single crystal neutron diffraction to determine the field/temperature dependence of the magnetic structure in YMn$_6$Sn$_6$. We find that the observed THE cannot arise from a magnetic skyrmion lattice, but instead from an in-plane field-induced double fan spin structure with *c*-axis components (DFC). Our work provides the experimental basis from which a microscopic theory can be established to understand the observed THE.

Two-dimensional (2D) magnetic kagome lattice materials, composed of corner sharing triangles and hexagons of magnetic ions separated by nonmagnetic buffer layers [Fig. 1(a)], are of great interest because they are candidates for quantum spin liquid [1], flat electronic bands [2-5], topological electronic [6], and magnetic behavior [7, 8]. For kagome metals [9], calculations using a simple tight-binding model with nearest-neighbor hopping reveal topologically protected linearly dispersive electronic bands near the Dirac point and dispersionless flat bands as confirmed by angle resolved photoemission experiments [2, 3, 5, 6]. On the other hand, the topological Hall effect (THE), a hallmark of the topological chiral spin textures termed magnetic skyrmions seen in field effect of non-centrosymmetric magnets [10, 11], has only recently been seen in antiferromagnetic kagome metal $YMn_6Sn_6$ near room temperature [Figs. 1(b-i)] [12]. The observation of THE requires a non-collinear spin texture, which has non-zero spin chirality [$\chi = \mathbf{S}_i \cdot (\mathbf{S}_j \times \mathbf{S}_k) \neq 0$, where $\mathbf{S}_i$, $\mathbf{S}_j$, $\mathbf{S}_k$ are the three nearest spins], that can induce non-zero Berry curvature acting as fictitious magnetic field for the conduction electrons to give rise to the THE [13-16]. For 3D itinerant-electron magnet MnSi, the skyrmion lattice is stabilized in the temperature regime at the border between paramagnetic and long-range helimagnetic phases perpendicular to a small applied magnetic field independent of the field direction to the atomic lattice [Figs. 1(j,k)] [17]. Since kagome metal such as $YMn_6Sn_6$ are highly 2D and the observed THE is strongly field directional dependent and becomes the largest for an applied magnetic field parallel to the Mn kagome plane [Fig. 1(i)] [12], it is unclear if the observed THE is also due to the formation of a field-induced magnetic skyrmion lattice. To understand the microscopic origin of THE in $YMn_6Sn_6$, it is therefore important to determine the magnetic structure of $YMn_6Sn_6$ as a function of applied field and its correlation with the observed THE.

$YMn_6Sn_6$ crystallizes in a hexagonal structure with space group *P6/mmm* and lattice parameters $a = b = 5.536$ Å, c = 9.019 Å [18, 19]. As shown in Fig. 1(a), the Mn atoms form kagome lattice slabs that stack along the *c*-axis alternating between Y-$Sn_3$ and included in its Mn-Sn1-Sn2-Sn1-Mn layers [Figs. 1(b,c)]. Based on powder neutron diffraction measurements, a helical antiferromagnetic (H-AFM) ground state has been reported [Figs. 1(b-d)] [19, 20]. While Mn spins in each kagome bilayer are collinear ferromagnetic (FM) and confined within the layer [Fig. 1(a)], they rotate from one bilayer to the next bilayer as shown in Figs. 1(c,d) and form helical stacking of the FM bilayers (termed "double-flat-spiral" magnetic structure) [21], as confirmed by the presence of incommensurate magnetic Bragg peaks in reciprocal space [Figs. 1(e-h)] [19, 22]. Previous magnetization and Hall resistivity measurements have shown that upon applying a magnetic field in the *ab*-plane (kagome layer) several distinct transitions occur, notably at 2.2 T, 6.8 T, and 9.8 T at 2 K [22, 23]. However, it is unclear what happens to the magnetic structure of the system, and the nature of the field-induced phase transition. Furthermore, THE have been observed in a distinct regime of the temperature and in-plane magnetic phase diagram with a maximum value near 240 K and 4 T [Fig. 1(i)] [12]. To determine the microscopic origin of THE, one must first determine the magnetic structure of $YMn_6Sn_6$ as a function of temperature and in-plane magnetic field.

In this paper, we use neutron diffraction to determine the temperature and in-plane magnetic field dependence of the spin structure in YMn$_6$Sn$_6$. At zero field and 2 K, we confirm the double-flat-spiral magnetic structure from the earlier work [19, 21]. Upon applying an in-plane magnetic field along the [1,0,0] direction, the system exhibits a first order magnetic phase transition for field above 2.2 T. Using small angle neutron scattering, we find no evidence of a skyrmion lattice in the temperature-field regime where the THE was observed. Surprisingly, our single crystal neutron diffraction measurements at 5 T and 2 K reveal that Mn spins on kagome lattice sites form an in-plane double fan structure [24] but with $c$-axis components, termed DFC (Figs. 2,3). Such a spin structure is different from a skyrmion lattice but satisfies the non-zero spin chirality $\chi = \mathbf{S}_i \cdot (\mathbf{S}_j \times \mathbf{S}_k) \neq 0$ requirement of the THE phase. Upon warming up 240 K, the in-plane field induced first order phase transition from the double-flat-spiral magnetic structure to the DFC structure is reduced to 1.8 T. In addition, the field-induced DFC magnetic structure has a large value of non-zero spin chirality at 3 T, and becomes a collinear ferromagnet for fields above 5.5 T. Our results are consistent with temperature-field phase diagram of the THE established from transport measurements [12], and therefore provided a basis from which a microscopic theory can be established.

Single-crystal neutron diffraction measurements were performed on the CORELLI elastic diffuse scattering spectrometer at the Spallation Neutron Source (SNS) [25], Oak Ridge Nation Laboratory (ORNL) and at the HB-3 thermal neutron triple-axis spectrometer at the High Flux Isotope Reactor (HFIR), ORNL. We define the momentum transfer **Q** in 3D reciprocal space in Å$^{-1}$ as $\mathbf{Q} = H\mathbf{a}^* + K\mathbf{b}^* + L\mathbf{c}^*$, where $H$, $K$, and $L$ are Miller indices and $\mathbf{a}^* = 2\pi(\mathbf{b} \times \mathbf{c})/[\mathbf{a} \cdot (\mathbf{b} \times \mathbf{c})]$, $\mathbf{b}^* = 2\pi(\mathbf{c} \times \mathbf{a})/[\mathbf{a} \cdot (\mathbf{b} \times \mathbf{c})]$, $\mathbf{c}^* = 2\pi(\mathbf{a} \times \mathbf{b})/[\mathbf{a} \cdot (\mathbf{b} \times \mathbf{c})]$ with $\mathbf{a} = a\hat{\mathbf{x}}$, $\mathbf{b} = a(\cos 120\, \hat{\mathbf{x}} + \sin 120\, \hat{\mathbf{y}})$, $\mathbf{c} = c\hat{\mathbf{z}}$ [Figs. 1(e-g)]. At CORELLI, full reciprocal space maps of the [$H$,0,$L$] and [$H$,$K$,0] scattering planes were obtained with up to 5 T applied magnetic field parallel and perpendicular to the kagome planes, respectively. Additional neutron diffraction was performed on HB-3 with a single crystal oriented in the [$H$,0,$L$] scattering plane with magnetic fields up to its maximum 8 T. By carrying out these neutron diffraction experiments, we were able to accurately determine the magnetic structure of YMn$_6$Sn$_6$ through a large area of temperature and magnetic field phase space.

We first use neutron diffraction to determine the zero-field magnetic structure of YMn$_6$Sn$_6$. Below $T_N$ = 333 K, magnetic Bragg peaks at an incommensurate wave vector $(0,0,\delta)$ away from nuclear Bragg peaks were observed [19, 20]. Figure 1(h) shows a map of reciprocal space in the [$H$,0,$L$] scattering plane, revealing a set of incommensurate peaks at $\mathbf{Q} = (H, 0, L \pm \delta), H, L = \pm 1, 2, \cdots$. We have refined the single crystal diffraction data using the software Jana2006 [26]. The intensity of the incommensurate peaks in the zero-field data indicates that YMn$_6$Sn$_6$ forms a double-flat-spiral magnetic structure as shown in Figs. 1(b-d) [21], including FM bilayers of the Mn kagome planes that couple antiferromagnetically to neighboring bilayers [8]. Every bilayer is rotated by $\delta d$ degrees in the plane, where $d$ is the distance between adjacent layers of moments aligned FM, and has no moment along the $c$-axis, as shown in Fig. 1(c). This is because the double-flat-spiral magnetic structure has a strong FM exchange interaction through Sn-Sn-Sn layers and a much weaker interaction through the Y-Sn layer [21]. The refinement results show that $\delta$ decreases upon cooling from $\delta = 0.261 \pm 0.008$ at 240 K to $\delta = 0.283 \pm 0.002$ at 2 K corresponding to a gradual change in helix rotation angle, $\delta d$. These results are consistent with earlier neutron diffraction work [19].

Next, the evolution of the magnetic phase is observed upon applying an in-plane magnetic field. As magnetic field is increasing, an abrupt change in phase occurs near B ~ 2.2 T at 2 K and B ~ 1.8 T at 240 K [Fig. 1(i)]. By performing both increasing and decreasing field sweeps of the $(0,0,2 - \delta)$ incommensurate peak, a first order phase transition is clearly shown in Fig. 2(a). Above this phase transition, a new double fan phase with a component along the *c*-axis (DFC) that emerges as a result of a spin-flop towards the direction of the applied field. This fan-like spin configuration is fully mapped by neutron diffraction measurements at 2 K, 5 T [Fig. 2(b)]. Due to interlayer weak AFM exchange coupling, after the spin-flop transition the spins have a *c*-axis canting [27]. This is shown in the magnetic field dependence of the incommensurate peaks along the $[H, 0, L]$ ($H = 1,2$) direction. Since neutron diffraction is only sensitive to the ordered moment component perpendicular to the wave vector $\boldsymbol{Q}$, one can use the wave vector/field dependence of the incommensurate peaks to determine the evolution of the double-flat-spiral magnetic structure above the critical field. For incommensurate peaks along the $[0,0, L \pm n\delta]$ ($n = 1,2$) direction, an in-plane field-induced ordered moment along the *c*-axis will not be revealed because it is along the wave vector direction [See $\boldsymbol{Q}_{(0,0,L)}$ in Fig. 1(h)]. On the other hand, for incommensurate peaks along the $[H, 0, L \pm n\delta]$ direction with $H = 1,2$ [See $\boldsymbol{Q}_{(2,0,L)}$ in Fig. 1(h)], neutron diffraction measurements would be sensitive to a field-induced *c*-axis component. Figures 2(c-d) show magnetic field dependence of the elastic scattering along the $[0,0, L]$ and $[H, 0, \pm n\delta]$ ($H = 1,2$) directions. From the field-induced magnetic incommensurate Bragg peaks in the insets of Figs. 2(d,e), we conclude that an in-plane magnetic field can actually induce a *c*-axis component in the new magnetic ordered state. To model the observed field-induced magnetic peaks, we consider a double fan structure with two sinusoidal functions representing the in-plane and out-of-plane components of the spin [Fig. 3(a)] [24]. The angle $\phi_{n,ab;n,c}$ of the ordered moment is $\phi_{n,ab} = \phi_{ab} \sin(n\, \delta d)$, and $\phi_{n,c} = \phi_c \sin(n\, \delta d - \varphi)$, where $\phi_{ab}$ and $\phi_c$ are the in-plane and *c*-axis fan amplitudes, respectively, and $\varphi$ is the phase differences between these two components. Using this model and our neutron diffraction data, we were able to refine the values of $\phi_{ab}$, $\phi_c$, $\varphi$, and $\delta d$ [Fig. 3(a)]. The result for 2 K, 5T is $\phi_{ab} = 0.24\pi$, $\phi_c = 0.26\pi$, $\varphi = 0.656\pi$, and $\delta d = 0.261\pi$ [Fig. 2(f)]. The resulting spin structure is shown in Fig. 3(a) along with its evolution with increasing applied magnetic field. This refinement includes many peaks (~50) and several have been included in Figs. 3(b-e) showing their response to magnetic field. In all cases, we find that the magnetic Bragg peaks are instrumental resolution limited, indicating a homogeneous DFC structure. This is clearly different from an intrinsically spin inhomogeneous skyrmion lattice structure [Fig. 1(j)] [10].

The DFC structure modeled represents a spin texture with non-zero spin chirality $\chi$. To see if a *c*-axis aligned magnetic field can also induce spin structure with non-zero spin chirality $\chi$, we carried out additional measurements at 2 K in the $[H,K,0]$ scattering plane with 0 and 5 T field along the *c*-axis and found no evidence of a spin structure with non-zero spin chirality (see supplementary information). This is clearly different from the skyrmion lattice expected to be field direction independent, but consistent with the field orientation dependence of the THE [12]. To further understand why the THE was clearly observed in YMn6Sn6 at 240 K and 4 T in-plane field, we carried out additional measurements at 240 K as a function of increasing field. Figures 4(b) and 4(c) show intensity maps of the [H,0,L] reciprocal space at 3 T and 5 T, respectively. At 3 T, we see incommensurate peaks at similar positions as that of the 5 T data at 2 K [Figs. 2(b) and 4(b)]. All incommensurate peaks are significantly suppressed at 5 T [Fig. 4(c)]. They are fully suppressed by ~6.5 T and YMn$_6$Sn$_6$ becomes spin polarized ferromagnet. Figure 4(d) shows magnetic field dependence of the incommensurate peaks, which reveals a phase transition around 1.8 T consistent with transport measurements [12]. The 3 T neutron diffraction data was fit to the DFC structure described previously. Due to less reflection being collected at this temperature and field, a cross check with the magnetization data [12] is necessary to determine the parameters. Our fitting results were that for 240 K, 3 T $\phi_{ab} = 0.455\pi$, $\phi_c = 0.379\pi$, $\varphi = 0.60\pi$, and $\delta d = 0.335\pi$ [Fig. 4(a)]. The refined values give calculated intensities in good agreement with the measured values as shown in Fig. 4(e). This refinement includes many peaks (~50), and several have been included in Figs. 3(f-g) showing their response to magnetic field. Thus, our refined magnetic structure using the DFC model supports the presence of a THE near 240 K, 3 T previously measured [12].

Since our neutron diffraction measurements suggest that the basic field-induced magnetic structures at 2 K and 240 K are similar, it is important to determine why the THE is larger at 240 K. Within the magnetic skyrmion lattice picture, the magnitude of THE is proportional to the ordinary Hall coefficient $R_0$ and the effective magnetic field (or non-spin spin chirality) [17]. From fitting the single crystal neutron diffraction data, we can completely determine the magnetic structures of the system at 2 K, 5 T and 240 K, 3 T. The non-zero c-axis component means that the spin configuration is no longer co-planer from layer to layer, giving rise to non-zero spin chirality, $\chi = \boldsymbol{S}_i \cdot (\boldsymbol{S}_j \times \boldsymbol{S}_k) \neq 0$, among the Mn spins along the c-axis. By comparing spin directions using the parameters from fitting the data, we can estimate spin chirality $\chi$ at 2 K, 5 T [Fig. 4(f)] and 240 K, 3 T [Fig. 4(g)]. Here the three spins $\boldsymbol{S}_i$, $\boldsymbol{S}_j$, and $\boldsymbol{S}_k$ are chosen from three consecutive unit cells. Averaging over 100 unit cells, the spin chirality at 2 K, 5 T is -0.726, which increases in magnitude to -0.864 at 240 K, 3 T. This, along with vanishingly small $R_0$ below 100 K [12], provides direct evidence for enhanced THE at 240 K and no THE at 2 K.

Our determination of the magnetic structures of YMn$_6$Sn$_6$ as a function of temperature and magnetic field provided the basis to understand the observed THE from the transport measurements [12]. In the unified molecular field theory (MFT) for an insulating helical antiferromagnet containing identical crystallographically-equivalent spins with weak c-axis magnetic exchange coupling [28], one would expect that an in-plane applied magnetic field should drive the helical structure into a fan structure without c-axis spin component [24]. A simple fan structure without c-axis spin component would result in zero spin chirality and therefore no THE. Our surprising discovery of an in-plane magnetic field applied on the metallic double-flat-spiral magnetic YMn$_6$Sn$_6$ inducing fan-like c-axis component is essential to understand the observed THE. Since YMn$_6$Sn$_6$ is a good metal [12], one would expect long-range magnetic exchange couplings including Ruderman-Kittel-Kasuya-Yosida (RKKY) interaction where magnetic interactions on localized Mn moments are mediated through the conduction electrons. From recent inelastic neutron scattering study of spin waves in YMn$_6$Sn$_6$, we see that the nearest neighbor magnetic exchange couplings within the kagome layer and in-between the bilayer along the c-axis are $J_1 \approx -28$ meV and $J_3 \approx -23$ meV, respectively [Fig. 1(a) and 1(c)] [8]. Since the FM magnetic exchange couplings within the bilayer $J_2$ is expected to be much larger than $J_3$, the comparable values of FM $J_1$ and $J_3$ suggest that there must be additional AFM exchange interactions along the $c$-axis to account for the in-plane field induced $c$-axis moment. Our work thus calls for a more detailed inelastic neutron scattering experiment to completely determine the magnetic exchange interactions in the system, particularly in the temperature and field regime where THE is observed.

In summary, we have used neutron diffraction to completely determine the magnetic structures of YMn$_6$Sn$_6$ as a function of temperature and magnetic field, focusing on the regime where the THE is observed. We show that the observed THE cannot be due to the skyrmion lattice typically associated with materials exhibiting THE. Instead, we discovered that an in-plane magnetic field can drive the system from a double-flat-spiral magnetic structure to a double fan spin structure with $c$-axis components. Such effect can only happen when there are long-range AFM exchange couplings along the $c$-axis, and the in-plane magnetic exchange couplings have similar magnitude as that of the $c$-axis magnetic exchange couplings. Our work provided the basis from which a future microscopic theory for the THE can be established.

During the process of preparing the present paper, we became aware of a preprint reporting magnetic field dependence of the transport, magnetic structures, and theory in YMn$_6$Sn$_6$ [29].

The neutron scattering work at Rice is supported by US NSF-DMR-1700081 and by the Robert A. Welch Foundation under Grant No. C-1839 (P.D.). H.C.L. was supported by the National Key R&D Program of China (Grants No. 2018YFE0202600, 2016YFA0300504), the National Natural Science Foundation of China (No. 11774423, 11822412), the Fundamental Research Funds for the Central Universities, and the Research Funds of Renmin University of China (RUC) (18XNLG14, 19XNLG17). A portion of this research used resources at the Spallation Neutron Source and the High Flux Isotope Reactor, a DOE Office of Science User Facility operated by ORNL.

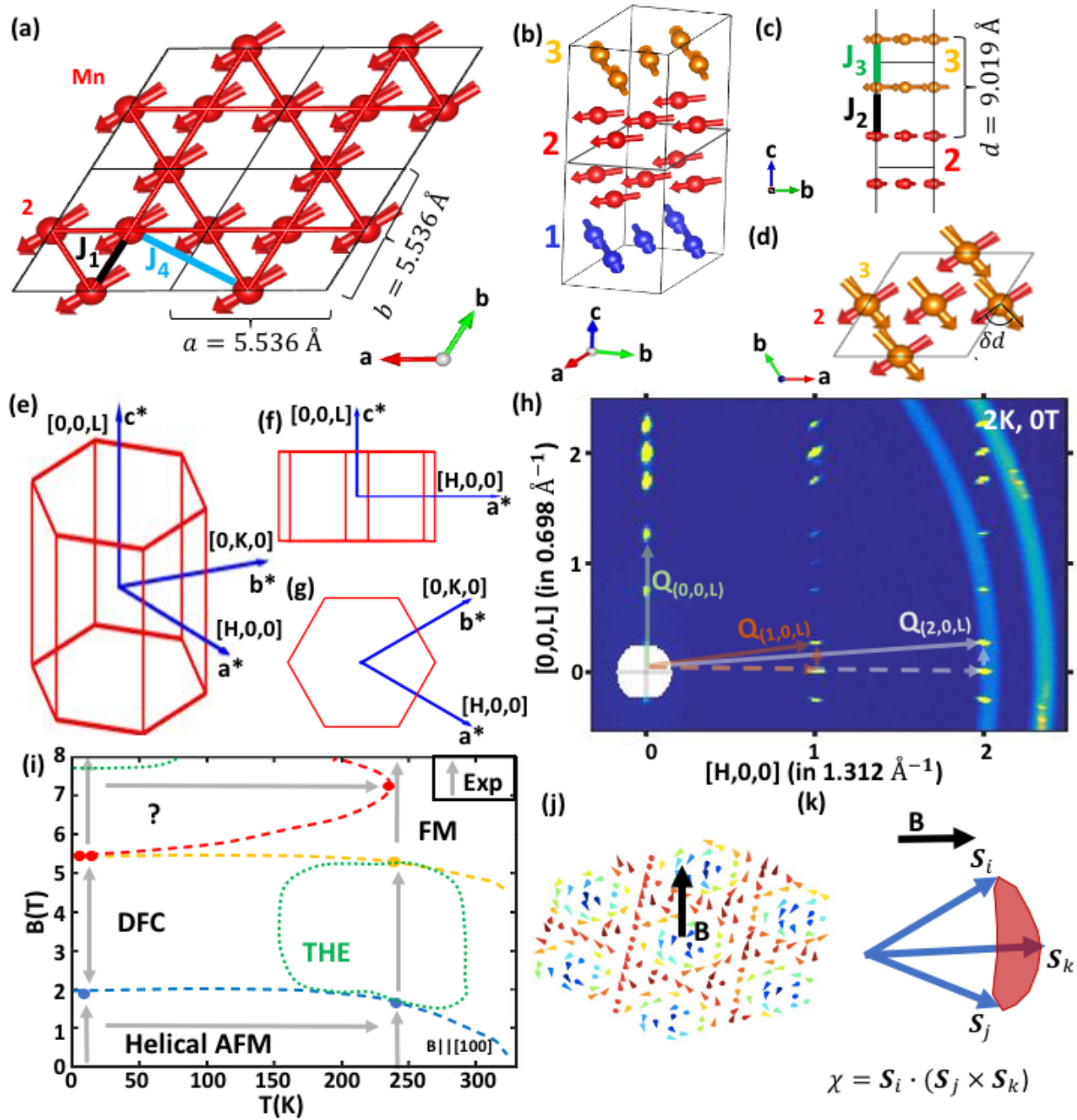

**Fig. 1.** a) Top view of the Mn bilayer in YMn$_6$Sn$_6$ crystal. Mn atoms form a kagome lattice with magnetic moments approximately collinear in each bilayer. The nearest neighbor and next-nearest neighbor magnetic exchange couplings within the Mn plane are $J_1$ and $J_4$, respectively. b) Zero-field magnetic structure of YMn$_6$Sn$_6$. c) Side view of structure shown in (b) where $d$ is the distance between kagome bilayers along the $c$-axis. The nearest neighbor magnetic exchange couplings along the $c$-axis are $J_2$ and $J_3$. The distance between collinear bilayers is 4.472 Å and the distance between non-collinear layers is 4.547 Å, corresponding to $J_2$ and $J_3$. d) Top view of structure shown in (b) where $\delta d$ is the rotation angle between kagome bilayers. e) Reciprocal space where [$H$,0,0], [0,$K$,0], and [0,0,$L$], or a*, b*, c* respectively, are specified. f) Reciprocal space in the [$H$,0,$L$] plane. g) Reciprocal space in the [$H$,$K$,0] plane. h) Single crystal neutron diffraction in the [$H$,0,$L$] plane at 2 K and 0 T. Incommensurate peaks are observed along

(0,0,*L*), (1,0,*L*), and (2,0,*L*) as indicated by Q$_{(0,0,L)}$, Q$_{(1,0,L)}$, and Q$_{(2,0,L)}$, respectively. i) Phase diagram of YMn$_6$Sn$_6$ with an in-plane magnetic field where a double-flat-spiral magnetic structure exists at fields below ~2 T. Between 2 and 5 T a DFC spin structure emerges. At low temperatures, above ~5.5 T a new unknown phase denoted "?" exists while at temperatures above ~240 K fields above ~5.5 T cause a forced FM state. Gray errors indicate where neutron diffraction measurements were taken. j) A schematic representation of a nanoscale skyrmion lattice with field applied perpendicular to the kagome lattice plane. k) A schematic illustrating the scalar spin chirality χ that is non-zero when the three spins $S_i$, $S_j$, and $S_k$ are non-coplanar.

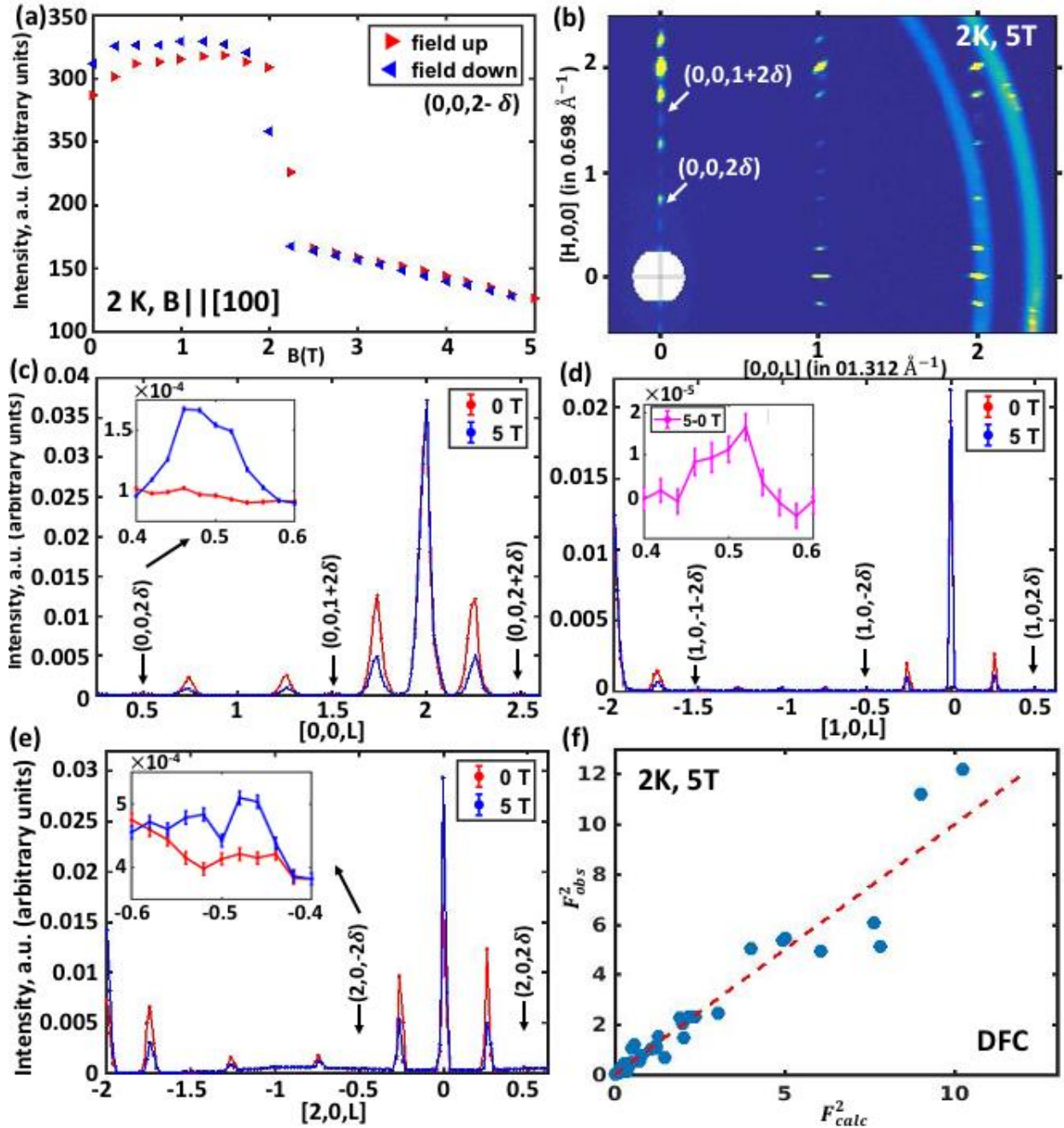

**Fig. 2.** a) The field dependence of the $(0,0,2-\delta)$ peak intensity at 2 K with the hysteresis observed between scanning fields up and down near 2 T. b) Single crystal neutron diffraction in the [$H,0,L$] plane of a YMn$_6$Sn$_6$ single crystal at 2 K and 5 T. Satellite peaks are observed at $(0,0,\pm 2\delta)$ and $(0,0,\pm\delta)$. c) Magnetic field dependence of [$0,0,L$] at 2 K. Inset shows $(0,0,2\delta)$ peak. d) Magnetic field dependence of [$1,0,L$] at 2 K. Inset shows the subtracted field-on, field-off peak at $(1,0,2\delta)$. e) Magnetic field dependence of [$2,0,L$] at 2 K. Inset shows $(2,0,-2\delta)$ peak. f) . The observed intensity plotted as a function of calculated intensity using DFC model with $c$-axis component. Red dashed line is the $y=x$ guideline. Overall $\chi^2 = 5.5$.

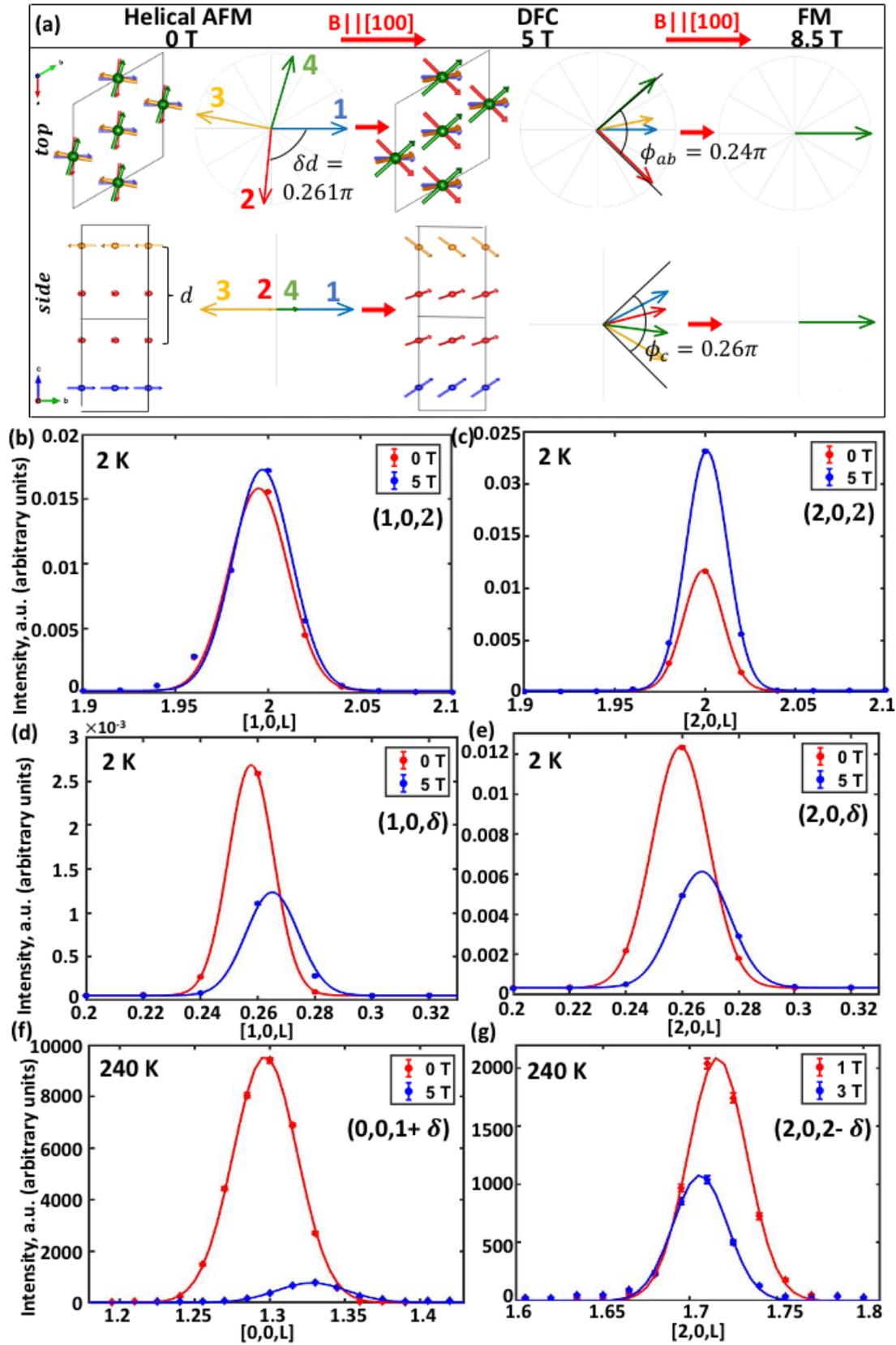

**Fig. 3.** a) The evolution of the magnetic structure at 2 K with applied field from the top and side views. Numbers 1-4 represent different Mn kagome bilayers stacked along the *c*-axis. Below 2 T, the helical AFM structure persists. An in-plane rotation angle of $\delta d = 0.261\pi$ exists between kagome bilayers while there is no *c*-axis canting. Above 2 T, the DFC spin structure emerges. At 5 T the in-plane fan spread is $\phi_{ab} = 0.24\pi$ and there is a *c*-axis canting with a spread of $\phi_c = 0.26\pi$. Increasing field above 8.5 T results in a forced FM state. Magnetic field dependence of b) the (1,0,2); c) the (2,0,2); d) the (1,0, $\delta$); e) the (2,0, $\delta$) peaks at 2 K. f) Magnetic field dependence of the (0,0, $1 + \delta$) satellite peak at 240 K. g) Magnetic field dependence of the (2,0, $2 - \delta$) satellite peak at 240 K.

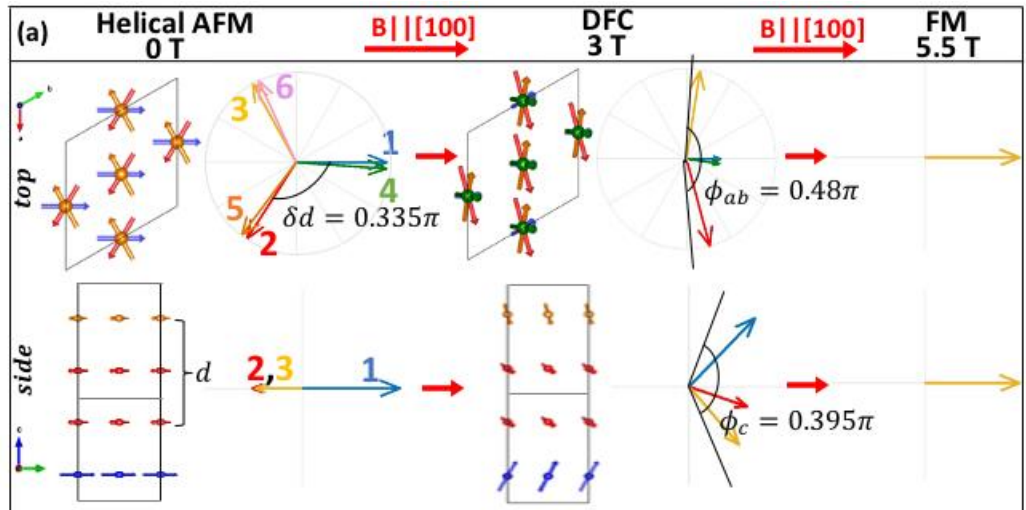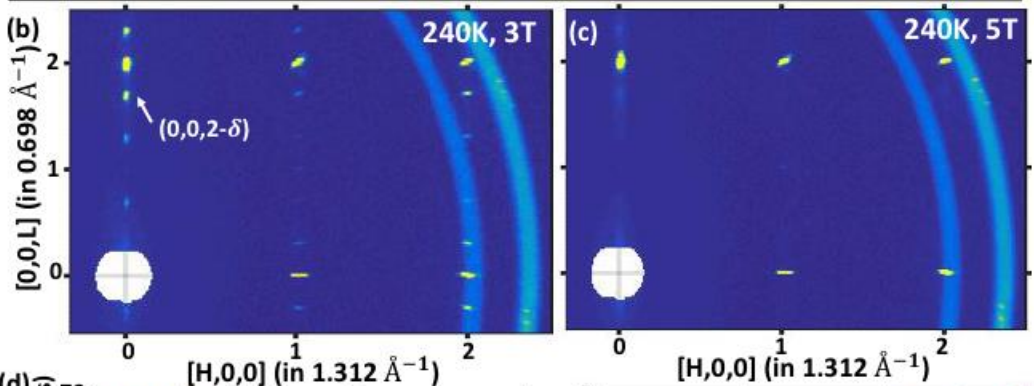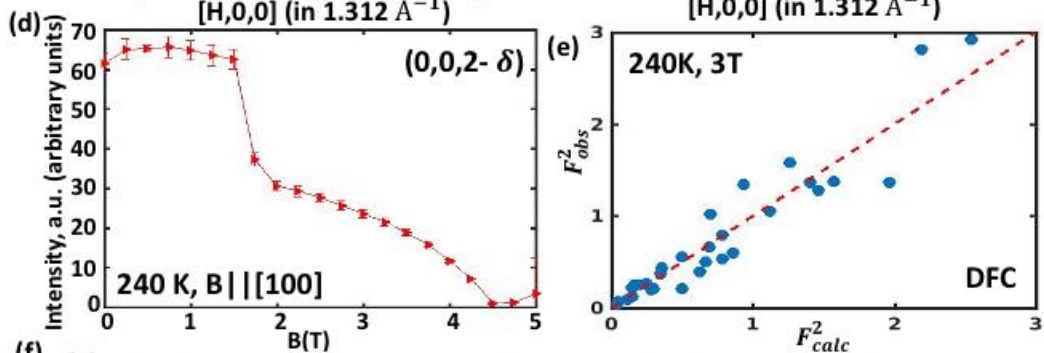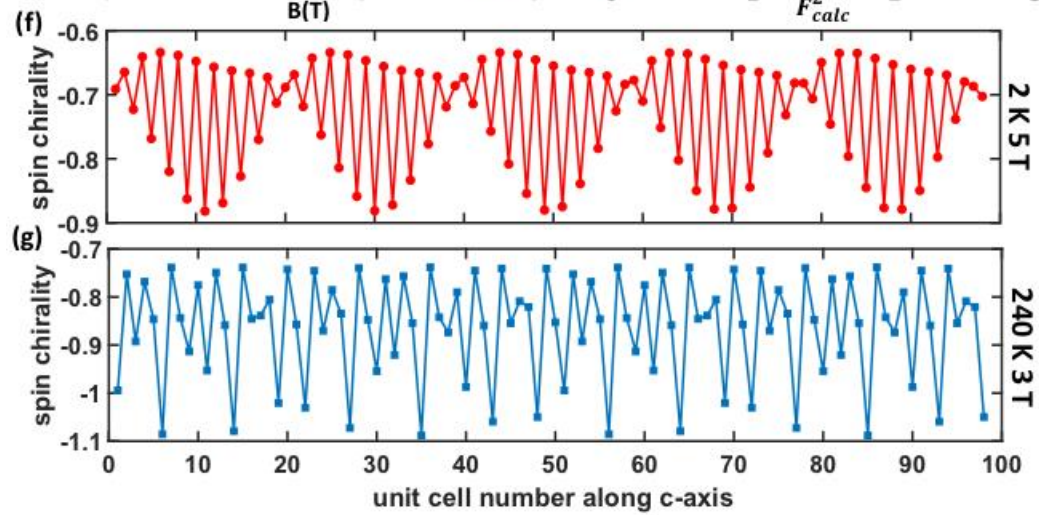

**Fig. 4.** a) The evolution of the magnetic structure at 240 K. Numbers 1-6 represent different Mn kagome bilayers stacked along the $c$-axis. Below 2 T, the helical AFM structure persists. An in-plane rotation angle of $\delta d = 0.335\pi$ exists between kagome bilayers while there is no c-axis canting. Above 2 T, the DFC spin structure emerges. At 3 T the in-plane fan spread is $\phi_{ab} = 0.48\pi$ and there is a $c$-axis canting with a spread of $\phi_c = 0.395\pi$. Increasing field above 5.5 T results in a forced FM state. b) Single crystal neutron diffraction in b) the [$H$,0,$L$] plane at 240 K and 3 T. c) in the [$H$,0,$L$] plane at 240 K and 5 T, where incommensurate peaks are nearly fully suppressed. d) Magnetic field dependence of the $(0,0,2-\delta)$ satellite peak at 240 K. e) The observed intensity plotted as a function of calculated intensity using fan-like model with AFM $c$-axis component. Red dashed line is the $y=x$ guideline. Overall $\chi^2 = 1.53$. The reduced $\chi^2$ is due to less reflection and reduced peak intensities comparing with 2 K, 5 T. Spin chirality $\chi$ as a function of unit cell number along the $c$-axis at f) 2 K, 5 T, and g) 240 K, 3 T.

*Supplemental materials:*

**SI1: Neutron diffraction Measurements**

The aim of this work was to connect the observed THE in YMn$_6$Sn$_6$ to its unknown magnetic structure. In several materials including MnSi [17] and Gd$_2$PdSi$_3$ [30], a skyrmion phase was observed that coincides with the measurement of the THE. Skyrmions stabilized by Dzyaloshinskii–Moriya interaction (DMI) typically range from 1–100 nm in size varying with material parameters, applied magnetic field, and temperature [30]. Considering this possibility of a skyrmion lattice in YMn$_6$Sn$_6$, small angle neutron scattering (SANS) was performed at the NG7 SANS instruments at the NIST Center for Neutron Research (NCNR). A 9 T horizontal field magnet was used with two orientations B‖[100] and B‖[001] through a temperature range of 5 to 300 K. Two configurations with sample to detector distances of 15.3 m and 2 m were used to span a total range of q from ~0.002 to 0.125 Å$^{-1}$. No skyrmion lattice signatures were observed under any temperature or field orientation throughout this q range. The absence of skyrmions is shown at 240 K, 3.5 T in the 15.3 m detector configuration in Fig. S1(a) and the 2 m detector configuration in Fig. S1(b). Instead, we determine the non-collinear spin texture of YMn$_6$Sn$_6$ with wide angle neutron measurements.

Single-crystal neutron diffraction measurements were performed on the CORELLI elastic diffuse scattering spectrometer at the Spallation Neutron Source (SNS), Oak Ridge Nation Laboratory (ORNL) and at the HB-3 thermal neutron triple-axis spectrometer at the High Flux Isotope Reactor (HFIR), ORNL. At CORELLI, a single crystalline YMn$_6$Sn$_6$ sample was mounted to an Al rod and oriented in the [$H$,0,$L$] and [$H$,$K$,0] scattering planes. The mounted samples are shown for the field in-plane along [1,0,0] orientation in Fig. S2(a) and field out-of-plane along [0,0,1] orientation in Fig. S2(b). A 5 T vertical field superconducting magnet with a top loading closed cycled refrigerator was used and covered temperatures between 2 and 240 K and magnetic fields up to its maximum of 5 T. Data was collected with a white incident beam while rotating the sample through a 360° range with a 1.5° step size for 3 minutes at each angle. In Fig. S3(a), the zero-field 2 K data for sample 1 is shown and the incommensurate satellite peaks corresponding to wavevector $\delta = 0.261 \pm 0.008$ are observed. In Fig. S3(b), a 5 T magnetic field has been applied to sample 1. A peak is observed at $\delta = 0.2653 \pm 0.0019$ and $2\delta$. Since $\delta$ is close to ¼, the $2\delta$ peaks overlap near (0,0,0.5) as shown in the inset of Fig. 2(c) in the main text. Measurements were repeated with a second single crystal sample. The results with sample 2 are consistent with sample 1 except for a slight change in incommensurability. This is shown in Fig. S3(c) where the zero-field 2 K data has incommensurate peaks corresponding to $\delta = 0.2653 \pm 0.0019$ and in Fig. S3(d) where the application of 5 T peaks results in peaks at $\delta$ and $2\delta$ with $\delta = 0.2799 \pm 0.0019$. In the [$H$,$K$,0] scattering plane, intensity is only observed at the Bragg peaks at 2 K, 0 T [Fig. S4(a)], 2 K, 5 T [Fig. S4(b)], and 240 K, 5 T [Fig. S4(c)]. Additional measurements were done to closely study the evolution of the incommensurate peak with applied field. With sample 1 in a fixed orientation of 16.5° and temperature of 2 K, field was swept in 0.25 T increments with 5 minutes per increment between 0 and 5 T as shown in Fig. 2(a) in the main text. This was also done at 240 K as shown in Fig. 4(d) in the main text.

Neutron diffraction measurements were also performed on HB-3 at ORNL. In Fig. S2(c), the single crystal mounted in the [$H$,0,$L$] scattering plane at HB-3 is shown. An 8 T vertical magnetic field with a top loading closed cycled refrigerator was used to cover temperatures between 20 and 240 K and magnetic fields up to its maximum 8 T. Measurements were taken with incident and out-going neutron energies of $E_i = E_f = 14.7$ meV. The measurements taken below 5 T are consistent with our results from CORELLI. Bragg peaks were resolution limited and fit to Gaussian peaks as shown in the main text in Figs. 3(b-g). Above 5 T, we observe a new peak at (0,0,1/2) that has a maximum intensity at 6.5 T as shown in Fig. S5 (a). The temperature dependence of this peak was also measured [Fig. S5(b)]. This indicates that a new phase exists in the high field and low temperature regime, but further measurements will be required to fully characterize it. This regime is shown in the main text in Fig. 1(i) and is denoted by "?".

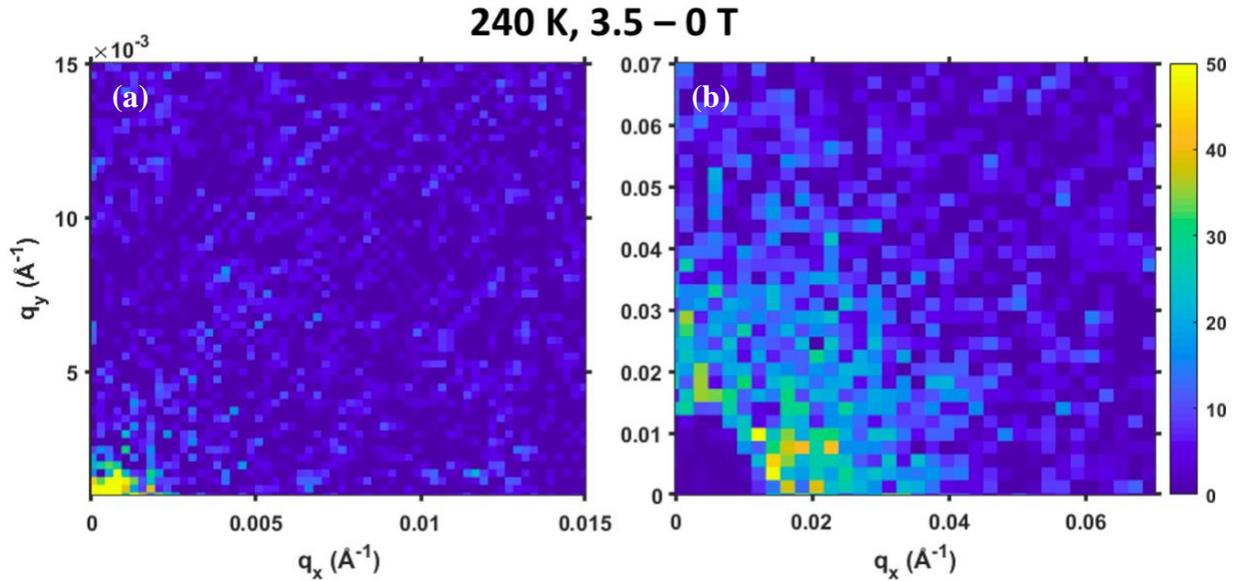

**Fig. S1.** Raw data of the SANS pattern of YMn$_6$Sn$_6$ at 240 K, 3.5-0 T at two different detector configurations, (a) 15.3 m and (b) 2 m, with different q ranges. Throughout these q ranges where skyrmions have been observed in other materials, YMn$_6$Sn$_6$ does not exhibit any skyrmion lattice signatures.

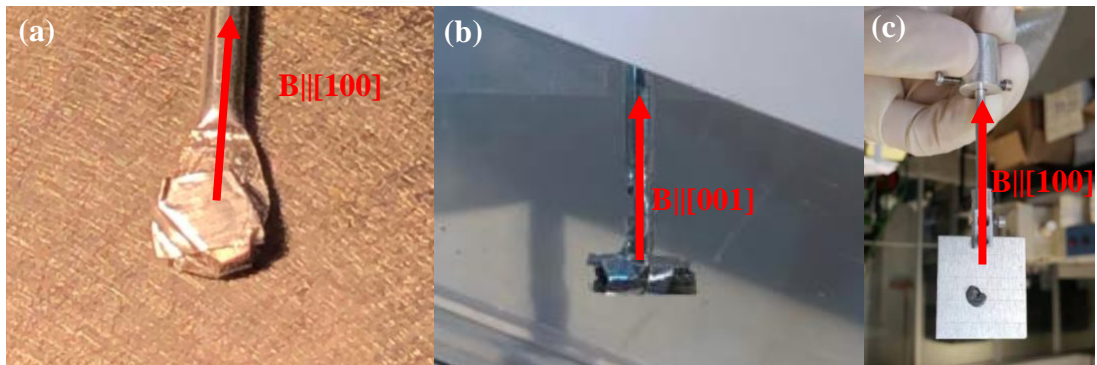

**Fig. S2.** YMn$_6$Sn$_6$ single crystals used in our neutron scattering experiments. (a) Single crystal affixed to Al rod with small amount of super glue and wrapped in aluminum foil for in-plane magnetic field, [H,0,L] scattering plane configuration at CORELLI. (b) Single crystal affixed to Al rod with small amount of super glue and wrapped in aluminum foil for c-axis magnetic field,

[*H*,*K*,0] scattering plane configuration at CORELLI. (c) Single crystal affixed to Al plate with CYTOP used in HB-3 triple axis experiment.

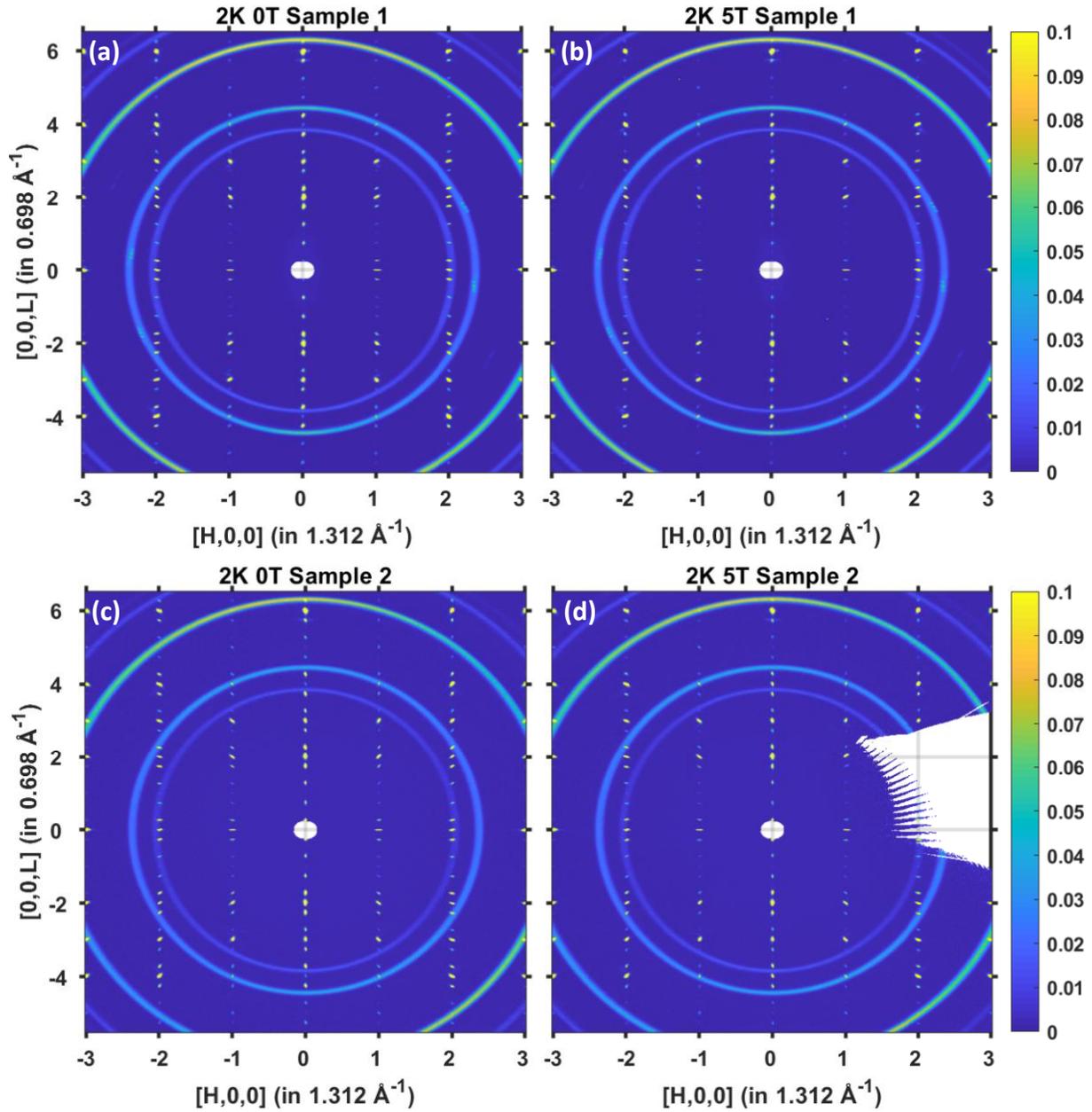

**Fig. S3.** Raw data of the diffuse neutron scattering pattern of YMn$_6$Sn$_6$ at 2 K, (a,c) 0 T and (b,d) 5 T for samples 1 (a,b) and 2 (c,d) obtained on CORELLI. The data were collected by rotating the sample up to 360 degrees along the vertical axis, with 1.5 degrees apart between orientations.

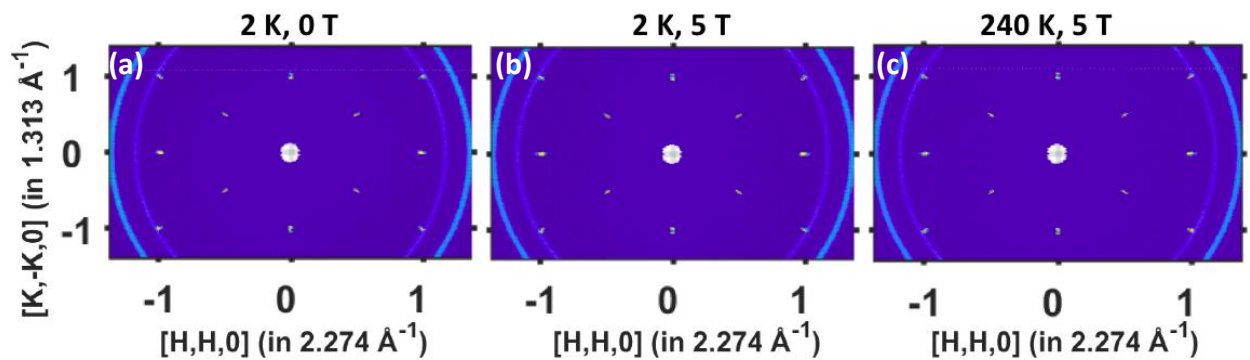

**Fig. S4.** Raw data of diffuse neutron scattering pattern of YMn$_6$Sn$_6$ in the [$H,K,0$] plane with field along the c-axis obtained on CORELLI at (a) 2 K, 0 T, (b) 2 K, 5 T, and (c) 240 K, 5 T.

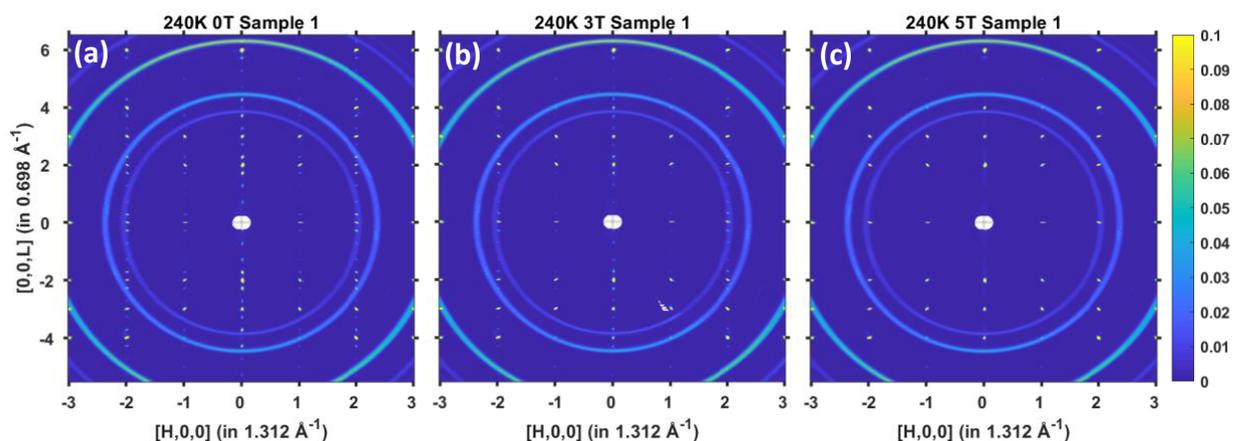

**Fig. S5.** Raw data of diffuse neutron scattering pattern of YMn$_6$Sn$_6$ at 240 K, (a) 0 T, (b) 3T, and (c) 5 T obtained on CORELLI. The data were collected by rotating the sample 360 degrees along the vertical axis, with 1.5 degrees apart between orientations.

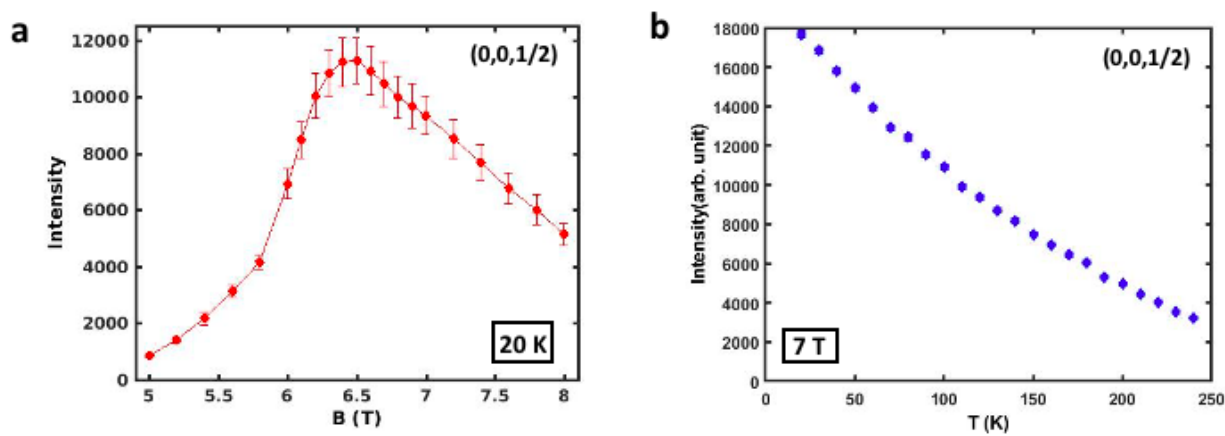

**Fig. S6.** At 20 K, neutron diffraction measurements observe a new peak above 5 T at (0,0,0.5). (a) The (0,0,0.5) peak reaches a maximum intensity at ~6.5 T and is afterwards suppressed with increasing magnetic field. (b) When temperature is increased, under 7 T the (0,0,0.5) peak is suppressed until ~240 K. This corresponds to an unknown magnetic phase denoted "?" in the phase diagram [Fig. 1(f)].

### SI2: Neutron Data Refinement

The Mantid package was used for data reduction [31]. The peak intensities for peaks in the [$H,0,L$] scattering plane were found by integrating along the [$H,0,0$] and [$0,K,0$] directions with $\Delta H=0.1$ and $\Delta K=0.1$ and fitting each peak to a Gaussian along [$0,0,L$]. The Gaussian fit also determined the position of the peak or wavevector $\delta$ as listed for each measured temperature, field, and sample in Table S1. Incommensurability varies slightly by sample and $\delta$ increases with increasing temperature. Using the determined peak intensities at each Bragg peak in zero-field, the FULLPROF software was used to refine the nuclear structure of YMn$_6$Sn$_6$ [32]. The result of this refinement is seen in Fig. S6. The zero-field magnetic refinements of CORELLI neutron diffraction data were carried out with the Jana2006 software for the helical antiferromagnetic (H-AFM) structure [26]. The non-centrosymmetric super-space group P6.1'(00g)-hs was used for the refinement of the incommensurate peaks in zero-field. The fitting was done at both 2 K and 240 K with an increased rotating angle $\delta d$ between layers at 240 K. The 2 K results are included in Fig. S8.

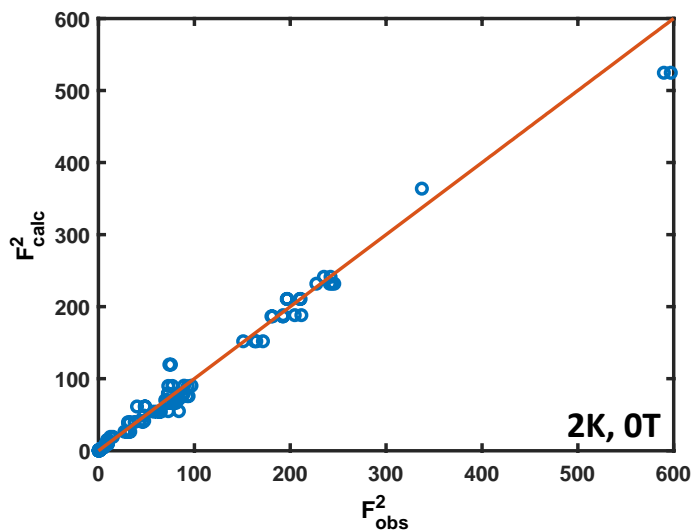

**Fig. S7.** The observed intensity plotted as a function of calculated intensity for the 2 K, 0 T nuclear refinement done using FullProf software. Red line is the y=x guideline. The number of reflections used was 128 and the overall chi^2 = 14.8.

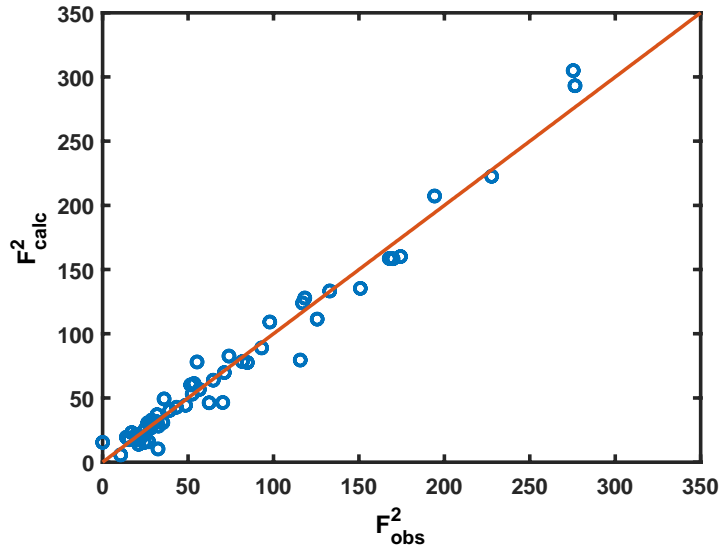

**Fig. S8.** The observed intensity plotted as a function of calculated intensity for the 2 K, 0 T magnetic refinement done using Jana2006 software. Red line is the y=x guideline. The number of reflections used was 125 and the overall chi^2 = 1.94.

## SI3: Least Square Fitting on the Incommensurate Peak Intensities with Magnetic Field

As demonstrated in the main text, the diffraction pattern of YMn$_6$Sn$_6$ under in-plane magnetic field developed a series of incommensurate peaks with non-zero $H$ indices, and therefore cannot be reproduced by the refined magnetic structure under zero magnetic field. Since magnetic neutron scattering function is proportional to the spin component that is perpendicular to the momentum transfer, $Q$, this strongly indicates that a non-zero $c$-axis component exists in the magnetic structure under field. A spin structure model with fan-like configuration in both $ab$-plane and $c$-axis has been considered. The model has three parameters, the in-plane spreading angle $\phi_{ab}$, the $c$-axis spreading angle $\phi_c$, and the phase difference $\varphi$ between these two components. The magnetic contribution to the elastic channel of the neutron scattering function $S(Q)$ can be calculate on a simulated lattice with 200 layers. The spins from each layer are considered to be colinear. To determine these parameters based on the observed incommensurate intensities, least square fittings on the calculated $S(Q)$ versus observed $S(Q)$ have been performed with a large set of parameter combinations on two data sets.

For the datasets collected at 2 K, 5 T, the incommensurability is 0.26, very close to the commensurate wave vector of 1/4. As a result, secondary satellite peaks from adjacent Bragg positions overlap at half-integer $L$ positions. This provides extra constrain over the fitting parameters. First, the spreading angle $\phi_{ab}$ was determined by fitting a dataset with 16 incommensurate peaks along the (0,0,$L$) cut, as shown in Fig S9(a). This is because the spreading of spins along the $c$-axis has little influences on the $H = 0$ reflections. The optimal fitting result was obtained with $\phi_{ab} = 0.24\pi$. Then a 2D iteration of the rest two parameters, $\phi_c$ and $\varphi$ was performed on the whole dataset of 40 reflections including (0,0,$L$), (1,0,$L$) and (2,0,$L$) cuts. The map of the inverse $\chi^2$ is plotted in Fig S9(b). The optimal parameter combination is

$\phi_c = 0.265\pi$, $\varphi = 0.656\pi$. The overall $\chi^2$ is 5.50 and the quality of the optimal fitting is shown in Fig. S8(c).

For the data set collected at 240 K, 3 T, the incommensurability is 0.31. The incommensurate peak intensities of this data set are 3 to 5 times weaker than the previous dataset, and no secondary satellite peaks can be identified. Therefore, the constrain from this dataset of 33 reflections from the (0,0,L), (1,0,L) and (2,0,L) cuts on the fitting parameters is not enough to determine an optimal combination. Instead, the overall $\chi^2$ can reach the optimal value of 1.53 if $\phi_{ab}$ and $\phi_c$ follows the curve shown in Fig S10(a) and $\varphi = 0.60\pi$. To further determine the parameter $\phi_{ab}$ and $\phi_c$, the total magnetization with different choices of $\phi_{ab}$ and $\varphi$ was calculated, assuming $\phi_{ab}$ and $\phi_c$ follow the curve in Fig S10(a). From magnetization measurement we know the total moment at 3 T at 240 K is in the range of 0.3 to 0.35, as shown with the red curves in Fig S10(b). Then the intersection of the $\varphi=108$ dashed line and the red curves gives the optimal parameter set to be $\phi_{ab} = 0.455\pi$, $\phi_c = 0.38\pi$, $\varphi = 0.60\pi$. The overall $\chi^2$ is 1.53 and the quality of the optimal fitting is shown in Fig. S9(c).

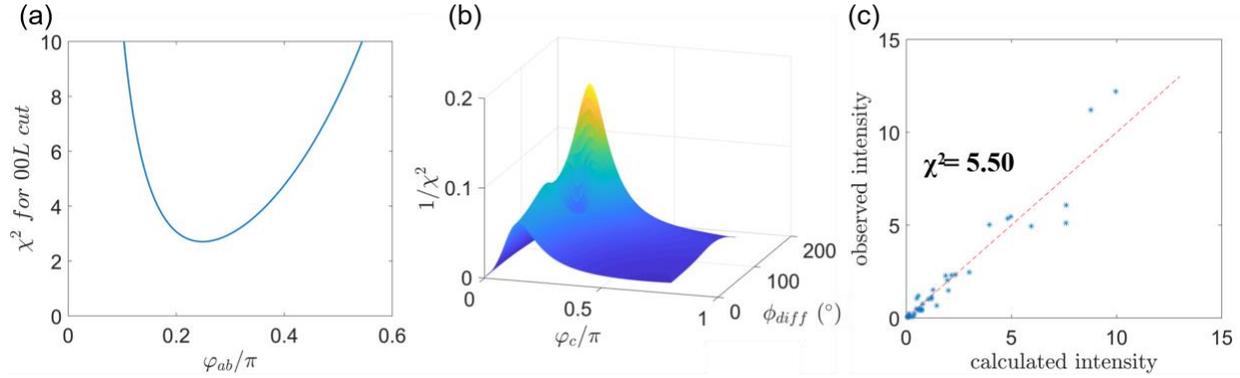

**Fig. S9.** Fitting results for 2 K, 5 T data set. Shown in (a) is the $\chi^2$ for (0,0,L) cut fitting to determine $\phi_{ab}$. The other two parameters, $\phi_c$ and $\varphi$ is not relevant for the $H = 0$ dataset. Shown in (b) is the map of $1/\chi^2$ as a function of $\phi_c$ and $\varphi$ while $\phi_{ab}$ is fixed at $0.24\pi$. The optimal parameter set is $\phi_{ab} = 0.24\pi$, $\phi_c = 0.265\pi$, $\varphi = 0.656\pi$. Shown in (c) is the observed intensity plotted as a function of calculated intensity. Red dashed line is the y=x guideline. Overall $\chi^2 = 5.5$.

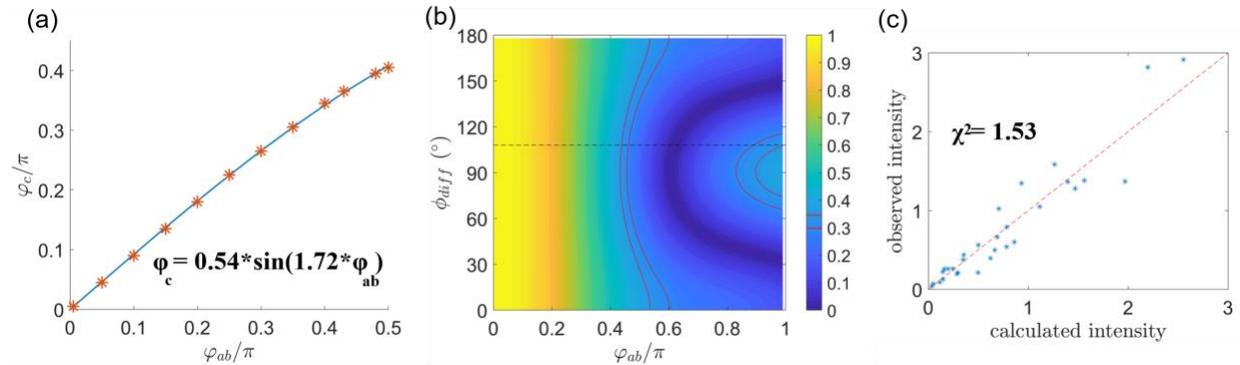

**Fig. S10.** Fitting results for 240 K, 3 T dataset. Shown in (a) is the curve of optimal $\phi_c$ as a function of $\phi_{ab}$. Shown in (b) is the contour map of total magnetization of the model as a function of $\phi_{ab}$ and $\varphi$, assuming $\phi_c$ follows the curve in (a). The magnetization measurement predicts the net moment at 240 K 3 T to be within the red curves. The optimal parameter set is determined to be $\phi_{ab} = 0.455\pi$, $\phi_c = 0.38\pi$, $\varphi = 0.60\pi$. Shown in (c) the observed intensity plotted as a function of calculated intensity. Red dashed line is the y=x guideline. Overall $\chi^2 = 1.53$.